\documentclass{article}
\usepackage{spconf,amsmath,graphicx,booktabs,url,xcolor}
\usepackage[T1]{fontenc}
\usepackage{mathptmx}
\usepackage{multirow}
\usepackage{svg}

\title{SideAge-RiskNet: Predicted Mammographic-Age Features for Future Breast Cancer Risk and Laterality Prediction}
\name{\shortstack{Haoyan Ding$^{1,2}$, Encheng Su$^{3}$, Chaowei Wu$^{4}$, Lixia Wang$^{2}$, Qian Wang$^{1,2}$, \\Garima Diwan$^{2}$, Linda Azab$^{2}$, Yibin Xie$^{2}$, Debiao Li$^{1,2}$}
\vspace{-4pt}
}

\address{
$^{1}$University of California, Los Angeles, Los Angeles, CA, USA \\
$^{2}$Biomedical Imaging Research Institute, Cedars-Sinai Medical Center, Los Angeles, CA, USA \\
$^{3}$Technical University of Munich, Munich, Germany \\
$^{4}$Siemens Medical Solutions USA, Inc., Malvern, PA, USA \\
}

\begin{document}
\ninept
\maketitle

\begin{abstract}
Whether breast-specific mammographic aging provides information beyond chronological age for future cancer prediction remains unclear. We propose SideAge-RiskNet, a framework integrating predicted mammographic-age features with breast-specific representations for multi-year breast cancer risk and conditional laterality prediction. A side-aware DenseNet121 trained on never-cancer reference examinations estimates left and right breast ages. Age-adjusted residuals and bilateral age descriptors are subsequently fused with corresponding image representations, while clinical variables contribute through an image--clinical risk branch. The framework jointly predicts cumulative cancer risk over 1--5 years and the subsequently affected breast among unilateral cancer cases. Across 4,480 EMBED and 248 external examinations, pre-cancer scans demonstrated pronounced bilateral age asymmetry. SideAge-RiskNet established state-of-the-art performance in 1--5-year risk assessment over Mirai and AsymMirai with mean risk AUROCs of 81.73\% on EMBED and 72.83\% on CSMC. Furthermore, it enabled conditional cancer laterality prediction, achieving AUROCs of 73.26\% and 82.53\%, respectively. This demonstrates the critical utility of bilateral aging phenotypes in localized breast cancer risk surveillance.
\end{abstract}

\section{Introduction}
\label{sec:intro}
Chronological age measures lifetime but does not fully characterize tissue condition, motivating biological-age markers such as epigenetic clocks\cite{horvath2013age}. Medical imaging offers a complementary, noninvasive approach to characterizing organ-specific aging\cite{ecker2026imaging,mueller2026imaging}. Deep learning models have estimated age from brain MRI \cite{cole2018brain}, chest radiographs \cite{raghu2021cxrage}, and retinal fundus images \cite{zhu2023retinal}, with discrepancies between predicted and chronological age associated with adverse health outcomes. Recently, Mammo-AGE demonstrated that mammographic age gaps are associated with breast cancer risk \cite{wang2025mammoage}. However, its age model training cohort was not strict which was without cancer-free check during longitudinal follow-up. Moreover, examination-level age prediction did not fully describe bilateral age differences and its cancer risk evaluation was limited to 2- and 5-year outcomes.

Prior studies have demonstrated the potential of mammography-based deep learning for future breast cancer risk prediction beyond conventional clinical risk models \cite{yala2019risk,arasu2023comparison,ellis2024risk}. Deep learning models such as Mirai \cite{yala2021mirai} have provided more comprehensive future breast cancer risk prediction up to 5 years from screening mammograms. AsymMirai further demonstrated the predictive value of bilateral mammographic asymmetry \cite{donnelly2024asymmirai}, consistent with earlier studies linking left–right tissue differences to near-term cancer risk \cite{tan2013asymmetry}. However, these approaches primarily provide patient-level risk estimates without identifying the breast in which cancer may subsequently develop. While individualized risk assessment already informs supplemental imaging recommendations\cite{monticciolo2023screening,lamb2026supplemental}, side-specific risk information could potentially support more localized surveillance and imaging assessment\cite{monticciolo2023screening}.
Moreover, whether mammographic aging differences between breasts provide complementary information for future cancer laterality remains unclear. This motivates investigating whether mammographic aging provides side-specific information about future cancer risk.

To address these gaps, we propose SideAge-RiskNet, a novel framework that integrates side-specific mammographic aging features into pre-diagnostic breast cancer assessment. Our contributions are threefold:
1. We introduce a side-aware aging network supervised on a rigorously verified, never-cancer reference cohort to estimate breast-specific biological ages, deriving bias-corrected descriptors of bilateral asymmetry and cancer-side divergence.
2. We fuse side-specific aging features with multi-view mammographic representations and clinical factors, achieving state-of-the-art cumulative 1--5-year risk prediction and outperforming established benchmarks including Mirai and AsymMirai.
3. We pioneer the task of pre-diagnostic breast cancer laterality prediction, demonstrating that biological age asymmetry provides vital, spatially specific signals that significantly improve affected-side discrimination.

\begin{figure*}[t]
\centering
\includegraphics[width=1\textwidth]{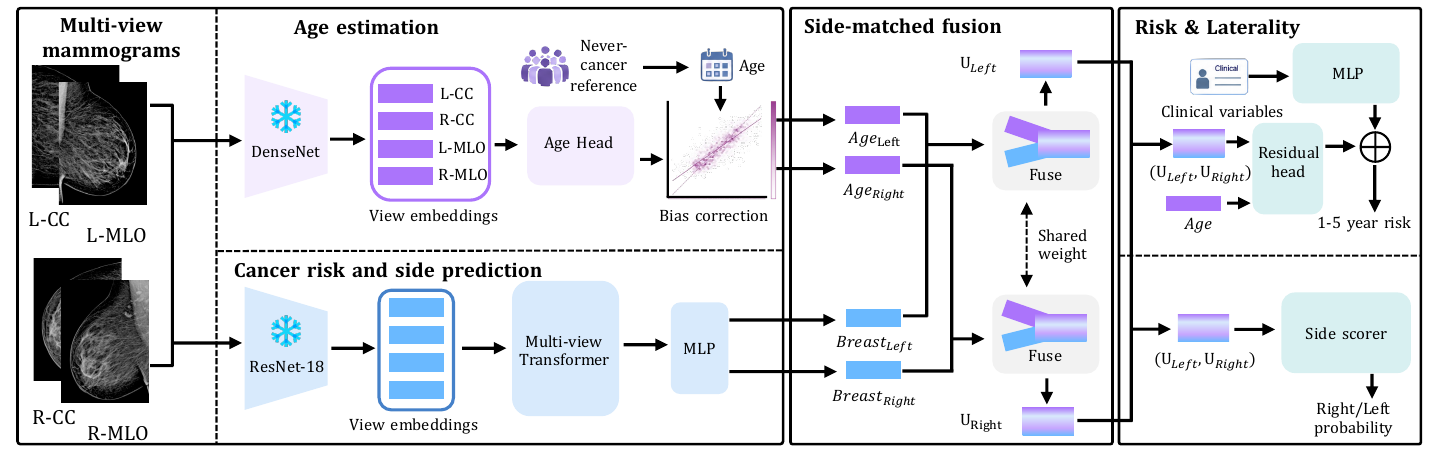}
\vspace{-13pt}
\caption{Overview of SideAge-RiskNet. Four-view mammograms are processed by separate age and risk branches. A DenseNet121 age model estimates breast-specific mammographic ages, from which bias-corrected age features are derived using a never-cancer reference.
A mammography-pretrained ResNet-18 and multi-view Transformer extract breast-level image representations.
Side-matched fusion modules with shared parameters combine image and predicted-age features into $\mathbf u_L$ and $\mathbf u_R$, supporting residual refinement of image--clinical 1--5-year risk predictions and conditional cancer laterality estimation.
}
\label{fig:framework}
\end{figure*}

\section{Methods}
\label{sec:methods}
\subsection{Problem setting and overview}
Our framework takes a four-view mammography examination—left and right craniocaudal (CC) and mediolateral oblique (MLO) views—together with clinical variables as input. As shown in Fig.~\ref{fig:framework}, we first estimate mammographic age, construct corresponding features. Then we integrate these related features into risk prediction model for both cumulative cancer risk within 1--5 years and the subsequently affected breast.

For age modeling, the reference cohort comprises
\emph{never-cancer} patients with no recorded breast cancer diagnosis throughout the available follow-up. For phenotype analysis, their examinations are compared with \emph{pre-cancer} examinations acquired before a subsequent breast cancer diagnosis. For risk prediction, labels are defined separately at each horizon $h\in\{1,2,3,4,5\}$ years. An examination is positive if breast cancer is diagnosed more than 30 days after imaging and within $h$ years, and negative if cancer-free follow-up is documented through that horizon.

\subsection{Side-specific mammographic age}
 A shared DenseNet121\cite{huang2017densenet} encodes all four views. Within each breast, learned attention weights aggregate CC and MLO features; a shared regression head predicts left and right ages, $\hat a_L$ and $\hat a_R$, and their mean is the examination age. Training uses chronological age from never-cancer reference examinations. Smooth-$L_1$ examination, side, and view losses are combined with bilateral and projection consistency penalties, with respective weights $1,1,0.25,0.20,0.05$. 
 The age model is fixed during risk-model training.

To account for age-related prediction bias
\cite{smith2019delta}, we remove the expected relationship between predicted and chronological age for each examination, breast, and projection aggregate $s$:
\begin{equation}
 r_s=\frac{\hat a_s-g_s(a)}{\sigma_s},
 \label{eq:age}
\end{equation}
where $g_s$ is a quadratic ridge regression fitted only on the EMBED age-training reference and $\sigma_s$ is its training residual standard deviation. For each breast, a six-dimensional age vector $\mathbf q_s$ contains its side residual, examination residual, absolute bilateral asymmetry, CC--MLO residual difference, and maximum and minimum side residuals. Features are standardized using risk-training statistics, which remain fixed during evaluation. For age-phenotype analysis, we additionally evaluate the linear bias-adjusted age gap (Linear BAG), expressed in years.

\begin{figure*}[t]
\centering
\includegraphics[width=0.8\textwidth]{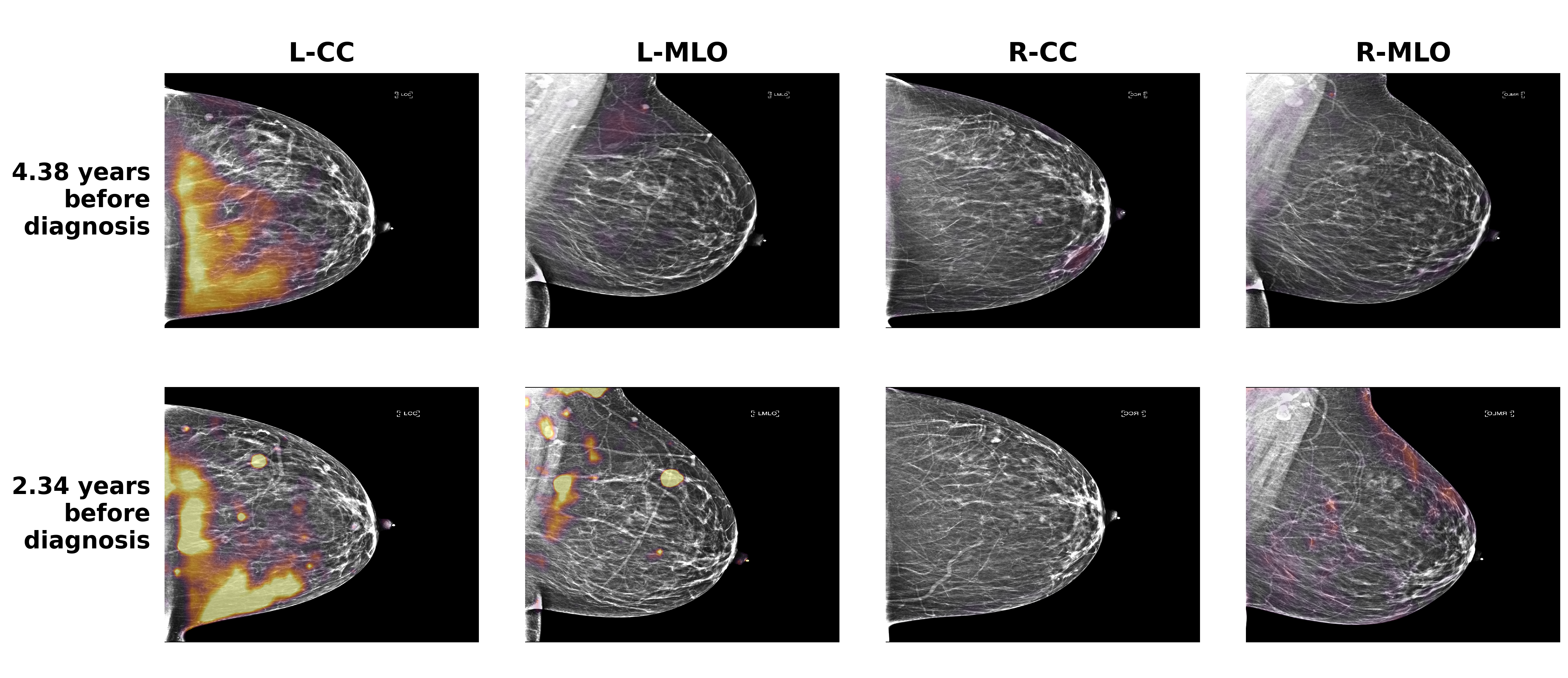}
\vspace{-15pt}
\caption{Grad-CAM attribution heatmaps of serial prediagnostic mammograms from the same held-out patient who subsequently developed left-sided breast cancer, shown for the predicted-age-enhanced model. Top and bottom rows correspond to examinations 4.38 and 2.34 years before diagnosis, respectively; columns show L-CC, L-MLO, R-CC, and R-MLO. Warmer colors indicate stronger positive image attribution to the model's output. The later examination shows more pronounced attribution in the future affected left breast.}
\label{fig:serial_case}
\end{figure*}

\subsection{Breast-level image--age fusion}
For risk prediction, a ResNet-18-based mammography encoder \cite{he2016resnet} initialized with Mirai risk-pretrained weights extracts a 512-dimensional representation from each view. The image encoder is partly frozen, while a trainable two-layer Transformer models relationships among the four views.

Let $\mathbf t_{s,CC}$ and $\mathbf t_{s,MLO}$ denote the contextualized tokens for breast $s\in\{L,R\}$. A shared breast-level MLP combines their appearance and projection differences:
\begin{equation}
\begin{aligned}
\mathbf b_s=\mathrm{MLP}_{\rm breast}\big(
[&\mathbf t_{s,CC},\mathbf t_{s,MLO},\\
 &|\mathbf t_{s,CC}-\mathbf t_{s,MLO}|,
 \mathbf t_{s,CC}-\mathbf t_{s,MLO}]
\big).
\end{aligned}
\end{equation}
The same MLP is applied to both breasts, producing 512-dimensional breast representations.

For the standardized age-feature vector $\mathbf q_s$ defined above, each component undergoes nonlinear basis expansion:
\begin{equation}
\begin{aligned}
c_{s,j}&=\operatorname{clip}(q_{s,j},-4,4),\\
\phi(q_{s,j})&=
\left[c_{s,j},c_{s,j}^{2},
\left(e^{-(c_{s,j}-\mu)^2/2}\right)_
{\mu\in\{-1.5,-0.5,0.5\}}\right].
\end{aligned}
\label{eq:agebasis}
\end{equation}

An age-feature MLP maps $\Phi(\mathbf q_s)$ to
$\mathbf z_s^{\rm age}$, which is concatenated with
$\mathbf b_s$ and processed by a fusion MLP:
\begin{equation}
\mathbf u_s=\mathrm{MLP}_{\rm fusion}
\big([\mathbf b_s,\mathbf z_s^{\rm age}]\big).
\label{eq:fusion}
\end{equation}
Each MLP shares its respective parameters across the left and right breasts. This module produces 
$\mathbf u_L$ and $\mathbf u_R$, integrating image information with anatomically corresponding predicted-age features. 

\subsection{Multi-horizon risk and laterality}
The breast-specific representations $\mathbf u_L$ and
$\mathbf u_R$ are aggregated into a patient representation:
\begin{equation}
\begin{aligned}
\mathbf u=[&
\mathbf u_L+\mathbf u_R,\,
|\mathbf u_L-\mathbf u_R|,\,
\mathbf u_L\odot\mathbf u_R,\,
\mathrm{MLP}_{\rm global}(\Phi(\mathbf q))].
\end{aligned}
\end{equation}
Here, $\odot$ denotes elementwise multiplication, and $\mathbf q$ is the global predicted-age feature vector. The bilateral terms encode shared appearance, asymmetry, and interactions.

A parallel image--clinical head combines global image
features $\mathbf g$ from the Transformer with clinical
variables $\mathbf c$ to estimate baseline risk logits:
\begin{equation}
\ell_h^0 = H_h(\mathbf g,\mathbf c),
\qquad h=1,\ldots,5.
\end{equation}
Here, $H$ includes clinical embedding, image--clinical
fusion, and risk prediction.

The patient representation $\mathbf u$ supplies complementary
image--age information through a residual risk adapter $D$:
\begin{equation}
\begin{aligned}
\delta_h&=0.5\tanh(D_h(\mathbf u)),\\
p_h&=\operatorname{sigmoid}(\ell_h^0+\delta_h).
\end{aligned}
\end{equation}
The adapter 
is followed by a five-output linear layer. Its final layer is initialized to zero, initially retaining the image--clinical risk predictions. The bounded corrections subsequently adjust the logits for the cumulative incident-cancer
targets. 

For laterality prediction, a shared scorer $S$ maps each
512-dimensional breast representation to a scalar score.
It consists of layer normalization, a 128-unit hidden layer,
GELU activation, dropout, and a linear output layer and gives the conditional right-side probability: 
\begin{equation}
 p_R=\operatorname{sigmoid}\big(S(u_R)-S(u_L)\big).
\end{equation}
Sharing the scorer applies the same comparison function to both
breasts. Laterality supervision uses only five-year-positive examinations with known unilateral malignant pathology. The side output is conditional left/right tendency, rather than calibrated absolute per-breast cancer risk.

The matched baseline retains the clinical inputs, breast-level
image processing, residual risk adapter, and laterality head,
but omits the predicted-age embeddings.

\subsection{Joint prediction objective}
For examination $i$ and horizon $h$, let $m_{ih}$ indicate sufficient outcome observation and $y_{ih}$ the cumulative cancer label. With training class weights $w_h=\min(N_h^-/N_h^+,100)$, the implemented risk objective is
\begin{equation}
\begin{aligned}
 \ell_{ih}&=-\kappa_h w_h y_{ih}\log p_{ih}
 -(1-y_{ih})\log(1-p_{ih}),\\
 \mathcal L_{\rm risk}&=\frac{\sum_{i,h}m_{ih}\ell_{ih}}
 {\max(1,\sum_{i,h}m_{ih})}.
\end{aligned}
\label{eq:riskloss}
\end{equation}
The positive-term horizon coefficients are $\kappa=(.25,1,1,1,1)$; negative terms are unscaled. Masking prevents insufficient follow-up from becoming a negative label. The five targets are cumulative events, not five independent biological outcomes. 

Let $\tau_i$ mask known unilateral five-year-positive cases, $y_i^R$ indicate right-side cancer, and $\omega_i$ be normalized inverse patient examination frequency. Then
\begin{equation}
\begin{aligned}
 \mathcal L_{\rm side}&=\frac{\sum_i\tau_i\omega_i\operatorname{BCE}(p_{R,i},y_i^R)}
 {\max(1,\sum_i\tau_i\omega_i)},\\
 \mathcal L_{\rm joint}&=\mathcal L_{\rm risk}+0.3\mathcal L_{\rm side}.
\end{aligned}
\label{eq:jointloss}
\end{equation}

\begin{table*}[t]
\centering
\normalsize
\caption{Risk and scan-level laterality AUROCs (\%). All methods share the same risk test. A dash indicates no evaluated laterality head. Bold and underlined values indicate the best and second-best results, respectively, within each dataset and column.}
\label{tab:main}
\setlength{\tabcolsep}{5pt}
\begin{tabular}{@{}llrrrrrrr@{}}
\toprule
Dataset & Model & 1y & 2y & 3y & 4y & 5y & 1--5y mean & Side AUROC \\
\midrule
\multirow{4}{*}{EMBED} & Mirai & 80.08 & 72.39 & 72.33 & 73.12 & 72.10 & 74.00 & -- \\
 & AsymMirai & 69.40 & 65.65 & 65.03 & 65.21 & 63.36 & 65.73 & -- \\
 & Our baseline & \textbf{94.43} & \underline{77.67} & \underline{75.47} & \underline{78.24} & \underline{82.36} & \underline{81.63} & \underline{71.88} \\
 & + predicted-age features & \underline{94.03} & \textbf{77.79} & \textbf{75.62} & \textbf{78.45} & \textbf{82.77} & \textbf{81.73} & \textbf{73.26} \\
\midrule
\multirow{4}{*}{CSMC} & Mirai & 69.48 & 63.28 & 62.30 & 64.89 & 63.69 & 64.73 & -- \\
 & AsymMirai & 67.05 & 60.13 & 55.90 & 57.30 & 53.12 & 58.70 & -- \\
 & Our baseline & \textbf{82.75} & \underline{71.15} & \underline{68.84} & \underline{72.71} & \textbf{67.63} & \underline{72.62} & \underline{78.75} \\
 & + predicted-age features & \underline{82.42} & \textbf{71.79} & \textbf{69.88} & \textbf{72.73} & \underline{67.34} & \textbf{72.83} & \textbf{82.53} \\
\bottomrule
\end{tabular}
\vspace{-0.6em}
\end{table*}

\section{Results}
\label{sec:results}
\subsection{Dataset and preprocessing}
We use the publicly released subset of the EMory BrEast imaging Dataset (EMBED) \cite{jeong2023embed}. After preprocessing and quality control, the age model is developed using never-cancer reference cohorts comprising 23,231 training, 2,880 validation, and 2,924 test examinations, with no patient overlap across these splits. A separate pre-cancer cohort containing 790 examinations is used for age phenotype analysis.
For downstream risk prediction, age-training patients are excluded from all risk partitions. The risk cohort comprises 12,887 training, 4,327 validation, and 4,480 test examinations, with patient-disjoint splits. 
External validation was performed using a retrospectively assembled cohort from Cedars-Sinai Medical Center (CSMC), comprising 248 examinations.

The input images are all preprocessed using an intensity scaling factor of 16. For CSMC, clinical variables are extracted from de-identified radiology reports using a large language model. The extracted variables are mapped to the same clinical categories used in EMBED to harmonize inputs across datasets.
\subsection{Implementation}
The age model is trained for 30 epochs using AdamW with an initial learning rate of $10^{-4}$, weight decay of $10^{-4}$. A cosine annealing schedule reduces the learning rate to $5\times10^{-6}$ over training.
The joint risk model is fine-tuned for 20 additional epochs using AdamW with a  weight decay of $10^{-4}$. The multi-view
Transformer and output normalization are updated at a learning rate of $10^{-5}$, while the breast-level fusion, predicted-age modules, residual risk adapter, and laterality head are trained at $3\times10^{-4}$. 

\subsection{Mammographic-age prediction and phenotypes}

Table~\ref{tab:age} summarizes age-estimation accuracy and scan-level age phenotypes. On the held-out never-cancer EMBED set, the side-aware model reduced MAE from 5.750 to 5.099 years ($p=0.0002$), with an RMSE of 6.585 years, Pearson $r=0.835$. CSMC MAE decreased from 8.495 to 7.682 years ($p=0.0140$), with an RMSE
of 9.872 years and Pearson $r=0.738$. It shows the importance of side-aware and better age estimation therefore transferred to CSMC.

We evaluated two age phenotypes: $\Delta_{\rm asym}$, the difference in mean absolute bilateral age asymmetry between pre-cancer and never-cancer examinations, and $\Delta_{\rm side}$, the mean paired difference between the future cancer side and the contralateral breast. Standardized residuals are expressed in SD units, whereas Linear BAG is expressed in years.

Pre-cancer examinations showed greater bilateral mammographic-age asymmetry in both datasets (Table~\ref{tab:age}). This difference reached  significance both in EMBED ($p=0.0478$) and CSMC ($p=0.0244$). Cancer-minus-contralateral differences were positive under both corrections in both datasets, which is significant in EMBED ($p=0.0384$). Although this difference did not reach statistical significance in the CSMC cohort—likely due to the smaller sample size—it demonstrated a consistent numerical trend toward an older cancer-side age. The results show greater pre-cancer asymmetry and positive cancer-side tendency under both corrections. 

\begin{table}[t]
\centering
\caption{Scan-level age results. MAE: original (Orig.) versus side-aware (Side), with reduction (Gain). $\Delta_{\rm asym}$: pre--never absolute L/R asymmetry difference; $\Delta_{\rm side}$: cancer--contralateral mean. SD/Years use residuals/Linear BAG.}
\label{tab:age}
\setlength{\tabcolsep}{2pt}
\begin{tabular}{lrrrrrrr}
\toprule
Dataset & \multicolumn{3}{c}{MAE (years)} & \multicolumn{2}{c}{$\Delta_{\rm asym}$} & \multicolumn{2}{c}{$\Delta_{\rm side}$} \\
\cmidrule(lr){2-4}\cmidrule(lr){5-6}\cmidrule(lr){7-8}
 & Orig. & Side & Gain & SD & Years & SD & Years \\
\midrule
EMBED & 5.750 & 5.099 & 0.651 & $+0.094$ & $+0.363$ & $+0.128$ & $+0.506$ \\
CSMC & 8.495 & 7.682 & 0.813 & $+0.309$ & $+1.168$ & $+0.042$ & $+0.111$ \\
\bottomrule
\end{tabular}
\vspace{-0.6em}
\end{table}

\subsection{Cancer risk and laterality prediction}
Table~\ref{tab:main} summarizes risk and laterality discrimination on EMBED and CSMC. Our baseline achieved higher risk AUROCs than the evaluated pretrained Mirai and AsymMirai checkpoints at all five horizons in both datasets. The results suggest stability across longer prediction horizons. Moreover, the model's advantage over both evaluated comparators persisted at every horizon in both cohorts, supporting the robustness of its comparative performance across prediction horizons and datasets. The incorporation of predicted-age features also consistently enhanced multi-year risk discrimination across both datasets. Unlike standard models such as Mirai and AsymMirai, SideAge-RiskNet extends risk assessment to predict conditional cancer laterality, highlighting the added value of side-specific phenotypic modeling. On both EMBED and CSMC, predicted-age features significantly improved patient-level laterality results, supporting further investigation of their spatially specific predictive information. 

Figure~\ref{fig:serial_case} presents Grad-CAM\cite{selvaraju2017gradcam} attribution heatmaps for two selected examinations from the same patient in the EMBED test cohort, acquired 4.38 and 2.34 years before a subsequent left-sided cancer diagnosis. The maps visualize image-branch contributions to the predicted-age-enhanced model's risk output. Compared with the earlier examination, the later examination shows stronger image-branch attribution around focal high-contrast structures and posterior tissue in the left breast. These observations illustrate variation in the model's spatial attribution across prediagnostic time points, which align closely with the growth trajectory of the lesion. Attribution is more focus on the future cancer side than on the contralateral breast, suggesting that the model places greater emphasis on image features from the subsequently affected side when estimating risk.

\section{Discussion and Conclusion}
\label{sec:discussion and conclusion}
This study connects mammographic-age estimation with future breast cancer risk and conditional laterality prediction. Side-aware age modeling improved estimation accuracy, and pre-cancer examinations showed greater bilateral age asymmetry. The incorporation of predicted-age features consistently enhanced mean risk AUROC and further improved patient-level laterality discrimination.

Several limitations warrant consideration. The small external cohort and only 18 first-year EMBED-positive examinations also limit precision. More external data validation is needed. The framework also requires separate age and risk image encoders, introducing additional inference. Future work will investigate shared-encoder architectures to reduce model size while retaining informative age representations for risk and laterality prediction.

In conclusion, SideAge-RiskNet integrates mammographic-age features with breast-specific image representations and models bilateral differences and interactions for multi-year risk and conditional laterality prediction. The findings motivate further investigation of bilateral aging phenotypes.

\vspace{-5pt}
\section{Compliance with Ethical Standards}
This retrospective study used publicly released EMBED
data and an institutional cohort
from Cedars-Sinai Medical Center. The analysis of the
institutional cohort was approved by the Cedars-Sinai
Institutional Review Board (protocol Pro00053107),
with a waiver of informed consent.

\section{Acknowledgments}
The authors declare no relevant conflicts of interest. 

\bibliographystyle{IEEEbib}
\bibliography{references}
\appendix
\end{document}